\documentclass[aps,twocolumn,amsmath,amssymb,preprintnumbers]{revtex4}
\usepackage{amsmath} \usepackage{amsfonts} \usepackage{amssymb}
\usepackage{bbm}
\usepackage{epsfig}
\usepackage{graphics}
\usepackage{graphicx}
\newcommand{\be}{\begin{equation}}
\newcommand{\ee}{\end{equation}}
\newcommand{\ba}{\begin{eqnarray}}
\newcommand{\ea}{\end{eqnarray}}
\newcommand{\nn}{\nonumber}
\newcommand{\kr}{\rangle}
\newcommand{\kl}{\langle}

\begin{document}

\title[ ]{Non-commutativity from coarse grained classical probabilities}

\author{C. Wetterich}
\affiliation{Institut  f\"ur Theoretische Physik\\
Universit\"at Heidelberg\\
Philosophenweg 16, D-69120 Heidelberg}

\begin{abstract}
Non-commutative quantum physics at the atom scale can arise from coarse graining of a classical statistical ensemble at the Planck scale. Position and momentum of an isolated particle are classical observables which remain computable in terms of the coarse grained information. However, the commuting classical product of position and momentum observables is no longer defined in the coarse grained system, which is therefore described by incomplete statistics. The microphysical classical statistical ensemble at the Planck scale admits an alternative non-commuting product structure for position and momentum observables which is compatible with the coarse graining. Measurement correlations for isolated atoms are based on this non-commutative product structure. We present an explicit example for these ideas. It also realizes the discreteness of the spin observable within a microphysical classical statistical ensemble.
\end{abstract}

\maketitle

The origin of the non-commuting product of operators is one of the old puzzles for the conceptual understanding of quantum physics. One may aim for an explanation of the axioms of quantum mechanics within a more general probabilistic formulation of the theory, based on an ensemble with positive probabilities and using the usual probabilistic relations between measurements, expectation values of observables and the probability distribution. It is often believed that such an approach must fail because of the commutativity of the classical correlation function for two observables. No go theorems using Bell's unequalities \cite{Bell} are based on the use of this classical correlation \cite{BS}. In this note we present an explicit example how a non-commuting product of observables arises from a classical statistical ensemble. This product is appropriate for the description of correlations of measurements in subsystems. Non-commutativity arises from ''coarse graining'' or a reduction of the available information. All features of quantum physics can be described in this way, including interference, tunneling, or the violation of Bell's inequalities. 

We use a classical statistical ensemble with states $\tau$ and positive probabilities $p_\tau\geq 0$ for the description of some microphysical system. As an example we take discrete variables at the Planck scale or beyond, which may be associated with occupation numbers for fermions. Within this setting we want to describe a single particle, such as an isolated atom, in terms of an appropriate subsystem. We assume that the observables relevant for the atom, i.e. position and momentum $X$ and $P$, are well defined as classical observables for the microphysical classical statistical ensemble. They have fixed values $X_\tau$ and $P_\tau$ for every state $\tau$ and their expectation values are computed according to the basic rules of classical statistics. Only a small part of the information available for the microphysical or ''Planck scale'' system will be needed and employed for the description of the atom at the ''Bohr scale''. Much of the information can be integrated out if we concentrate on the ''coarse grained'' subsystem and the details of the microphysical system will not be important for our setting. The concept of an atom as a subsystem is familiar in quantum field theory, where atoms are viewed as excitations of a complicated vacuum, which involves infinitely many degrees of freedom and collective phenomena such as spontaneous symmetry breaking. 

In the microphysical ensemble the classical product of position and momentum observables is well defined and commutative. For every state $\tau$ the value of the observable $X\cdot P=P\cdot X$ is given by $(X\cdot P)_\tau=(P\cdot X)_\tau=X_\tau P_\tau$. For our construction of the subsystem, however, this classical product remains no longer defined. The expectation value of the classical product $\kl X\cdot P\kr$ can be computed from the information available for the microphysical system, i.e. from the probability distribution $\{p_\tau\}$ and the values $X_\tau$ and $P_\tau$ for every state $\tau$. This complete statistical information is no longer available for the subsystem, since a large part of the microphysical information is lost by the coarse graining. The subsystem describing the atom is characterized by incomplete statistics \cite{3} for which the expectation value of $X\cdot P$ is no longer computable. 

On the other hand, the microphysical ensemble admits also a product structure for the observables $X$ and $P$ that is different from the classical product. This product is not commutative. It is this non-commutative product that remains computable in terms of the information available for the subsystem. We will argue that the outcome of good measurements of subsystem properties should be computable in terms of the statistical information available for the subsystem - otherwise one would measure ''environment properties'' beyond the properties of the isolated atom. Correlations between the results of such good measurements are described by the non-commutative product. We will show how this non-commutative product can be associated with the product of quantum  operators for position and momentum. 

The general concepts how incomplete statistics for subsystems leads to measurement correlations based on a non-commutative product of observables have been discussed in ref. \cite{CWAA,CWPO}. In ref. \cite{CWAA} it is shown in detail that these correlations indeed violate Bell's inequalities and are not in contradiction with the Kochen-Specker theorem \cite{KS}. What has been missing so far is a concrete example for a construction where a subsystem as an atom is described in terms of a classical ensemble for microphysical degrees of freedom. This note discusses such an example, where the microphysical degrees of freedom may be associated in a vague sense to fermions at the Planck scale. Our aim is not in any way the construction of a realistic particle physics model, and the model is only meant as an intuitive setting for a microphysical ensemble. The notions of ''Planck scale'' or ''Bohr scale'' can be replaced by arbitrary other scales, and an interpretation of the microphysical degrees of freedom in terms of fermions is not necessary. Our conceptual setting applies to a large class of possible fundamental particle physics models - the important notions involve only the coarse graining to a subsystem describing a one particle state. In principle, it can apply to arbitrary classical statistical systems provided that the coarse grained ''one particle state'' can be appropriately defined. In this way quantum features may be discovered in macroscopic classical statistical systems.

Usually, fundamental theories of particle physics are not based on a classical statistical ensemble. This holds despite the striking observation that the analytic continuation of the functional integral to euclidean space often allows for such a classical statistical formulation, as well known for lattice gauge theories. It is also well known that non-commuting operator structures can be defined for the classical statistical ensemble in euclidean space. Still, because of no go theorems based on Bell's inequalities, it is almost general belief that a quantum field theory in Minkowski space cannot be described by a classical statistical ensemble. The present note demonstrates that this belief is unfounded - the road to a classical probabilistic formulation of a fundamental theory of particle physics is open. 

In a second part of this note we implement another important property of quantum physics within a classical statistical description, namely that individual measurements of an observable only find eigenvalues of the associated quantum operator. We show how this quantum postulate follows from the standard classical statistical setting for measurements for two examples of quantum operators with a discrete spectrum, namely local projectors and the spin of a particle. 

\medskip\noindent
{\em  Function observables}

Let us start with a discrete chain of $L$ points, labeled by $s=1\dots L$, which may be placed on a circle. To each point we attach an ''occupation number'' $n_s$ which can take the values one or zero, $n_s=0,1$. We can associate this system with fermions on a one-dimensional lattice. Alternatively, one may interprete the $n_s$ as $L$ bits of a computer. A state of the system is given by an ordered sequence of $L$ bits, $\tau=\{n_s\}$, such that the number of different states $\tau$ equals $N_s=2^L$. We consider a classical statistical ensemble with positive and normalized probabilities $p_\tau$ for each state,
\be\label{1}
p_\tau\geq 0~,~\sum_\tau p_\tau=1.
\ee
Classical observables take a fixed value $A_\tau$ for every state $\tau$ and obey the standard definition for expectation values
\be\label{2}
\kl A\kr =\sum_\tau p_\tau A_\tau.
\ee

We will be interested in classical observables that can be associated to functions on a circle $f(x)$. For such an observable we define for each sequence $\{n_s\}=\tau$ a fixed function $f_\tau(x)$, which we normalize according to 
\be\label{2A}
\int dx f^2_\tau(x)=1.
\ee
The expectation value of such a ''function-observable'' reads
\be\label{3}
\kl f(x)\kr=\sum_\tau p_\tau f_\tau (x).
\ee
We can also define observables which measure properties of a function, like the ''roughness''
\be\label{4}
R_\tau=\int dx\big (\partial_x f_\tau(x)\big)^2
\ee
or a position variable
\be\label{5}
X_\tau=\int dx xf^2_\tau(x).
\ee
For these classical observables the expectation values are always given by eq. \eqref{2}. Intuively, one may realize such a ''function observable'' by associating a given range in $s$ to an interval in $x$, and taking $f_\tau(x)$ positive if there are more occupied $(n_s=1)$ than empty   $(n_s=0)$  bits in the interval, negative if there are more empty bits, and $f_\tau(x)$ close to zero for approximately equal numbers of occupied and empty bits in the interval. 

A simple particular function-observable can be obtained by the following construction. We choose a discrete set of $P$ equidistant points $x_i$ on the circle, with ''lattice distance'' $\epsilon$ and intervals $I(x_i)$ given by $x_i-\frac{\epsilon}{2}\leq x<x_i+\frac\epsilon 2$, such that these intervals cover the whole circle and $\int dx=P\epsilon$. We are interested in situations where the number of bits $L$ is much larger than $P$. To every bit $s$ we associate $x=x_0+s\delta$, with $\delta=(P/L)\epsilon$, and arbitrary $x_0$, e.g. $x_0=0$. Thus every bit $s$ ''belongs'' to one given interval $I(x_i)$. The function $f_\tau(x_i)$ is defined as
\be\label{6}
f_\tau(x_i)={\cal N}^{-\frac12}\sum_{s\in I(x_i)}(2n_s-1),
\ee
where the sum is taken over all $s$ which belong to the interval $I(x_i)$. Thus ${\cal N}^{1/2}f_\tau(x_i)$ is simply the number of occupied bits minus the number of empty bits within the interval $I(x_i)$. The normalization factor ${\cal N}$ is defined by the requirement
\be\label{7}
\int dx f^2_\tau(x)=\sum_i f^2_\tau(x_i)=1,
\ee
or
\be\label{8}
{\cal N}=\sum_{x_i}\big (\sum_{s\in I(x_i)}(2n_s-1)\big)^2.
\ee
(For simplicity we may take $L$ odd such that ${\cal N}>0)$ is guaranteed.) At the end we can take the continuum limit $\epsilon\to 0$ or $P\to\infty$ in the standard way, with $\int dx=\epsilon\sum_{x_i},f_\tau(x)=\epsilon^{-\frac12}f_\tau(x_i)$, either with fixed $\delta/\epsilon\ll 1$ or with $\delta/\epsilon\to 0$. 

The detailed description how we associate to a sequence $\tau$ a function $f_\tau(x)$ will not be important. As an alternative to the ''stepwise function'' \eqref{6} we may associate to each bit sequence $\tau$ a continuous and differentiable function even for finite $L$. This could be achieved by some smoothening prescription of the stepwise function. One could also use the complete system of periodic functions on the torus
\be\label{8A}
f(x)=\sum^{\infty}_{k=0}\left\{
a_k\cos\left(\frac{2\pi kx}{l}\right)+b_k\sin 
\left(\frac{2\pi kx}{l}\right)\right\}
\ee
and define an appropriate map from $\tau$ to a sequence of numbers $\{a_k,b_k\}$. For our purpose we only need the existence of a well defined map $\tau\to f_\tau(x)$. 

Our setting is easily generalized to more than one dimension, where $x^\mu$ may be coordinates on a $d$-dimensional torus. We can also consider more than one ''species'' of bits with occupation numbers $n_{\alpha,s}$ for every $s$ and $\alpha=1\dots F$ labeling different species. There are then $N_s=2^{FL}$ different states $\tau=\{n_{\alpha,s}\}$. The function observables $f_{\alpha,\tau}(x)$ can be defined separately for each species, for example by using in eq. \eqref{6} $n_{\alpha,s}$, and adapting the normalization according to
\be\label{9}
\sum_\alpha\int dx f^2_{\alpha,\tau}(x)=1.
\ee

For two species we can define a derivative observable
\be\label{10}
P_\tau=\int dx\big[f_{1,\tau}(x)\partial_xf_{2,\tau}(x)-
f_{2,\tau}(x)\partial_x f_{1,\tau}(x)\big].
\ee
We may also introduce a complex structure with a complex function $f_\tau(x)=f_{1,\tau}(x)+if_{2,\tau}(x)$, such that 
\be\label{11}
P_\tau=\int dx f^*_\tau(x)(-i\partial_x)f_\tau(x)~,~
\int dx f^*_\tau(x)f_\tau(x)=1.
\ee
The derivative observable resembles the momentum observable in quantum mechanics. Correspondingly, the position observable reads
\be\label{12}
X_\tau=\int dx f^*_\tau(x)xf_\tau(x).
\ee
Despite the similarities with the quantum formalism we recall that both $X_\tau$ and $P_\tau$ take sharp values in every state $\tau$. The classical correlation function is commutative,
\be\label{13}
\kl X\cdot P\kr_{cl}=\kl P\cdot X\kr_{cl}=
\sum_\tau p_\tau X_\tau P_\tau.
\ee

\medskip\noindent
{\em Coarse graining}

For the computation of the expectation values of position $\kl X\kr=\sum_\tau p_\tau X_\tau$ and momentum $\kl P\kr=\sum_\tau p_\tau P_\tau$ one needs much less information than contained in the probability distribution $\{p_\tau\}$ for the classical statistical ensemble. A specification of $\{p_\tau\}$ requires $2^{2L}$ real numbers. Many different $\{p_\tau\}$ lead to the same $\kl X\kr$ and $\kl P\kr$. In order to concentrate on the relevant information for a particle subsystem we perform a ''coarse graining'', as defined by the density matrix
\be\label{14}
\rho(x,x')=\sum_\tau p_\tau f_\tau (x) f^*_\tau(x').
\ee
As in quantum mechanics, the density matrix is normalized and hermitean
\be\label{15}
\text{Tr} \rho=\int dx \rho(x,x)=1~,~\rho^*(x,x')=\rho(x',x).
\ee
We can associate to the observables $X$ and $P$ the operators $\hat X$ and $\hat P$, 
\be\label{16}
\hat X(x',x)=\delta(x'-x)x~,~\hat P(x',x)=-i\delta(x'-x)\frac{\partial}{\partial x},
\ee
such that the expectation values obey
\ba\label{17}
\kl X\kr&=&\sum_\tau p_\tau X_\tau=\text{Tr}(\hat X\rho)=\int dx x\rho(x,x),\nn\\
\kl P\kr&=&\sum_\tau p_\tau P_\tau=\text{Tr}(\hat P\rho)\\
&=&-i\int dx'dx\delta(x'-x)\partial_x\rho(x,x').\nn
\ea
\\
The information contained in $\rho$ is therefore sufficient for the computation of $\kl X\kr$ and $\kl P\kr$.

The density matrix is not a property of a single state $\tau$ but rather involves the whole probability distribution $\{p_\tau\}$. In other words, the coarse graining corresponds to a map $\{p_\tau\}\to\rho(x,x')$. This map is not invertible - many different $\{p_\tau\}$ are mapped to the same $\rho(x,x')$ such that a large part of the microphysical information is lost by the coarse graining. We observe that eq. \eqref{17} constitutes precisely the quantum rule for the computation of expectation values of position and momentum from a quantum density matrix. Indeed, $\rho(x,x')$ is a positive matrix, such that together with eq. \eqref{15} it obeys all the required properties of the density matrix in quantum mechanics. The positivity of $\rho$, i.e. $\int_{x,x'}g^*(x)\rho(x,x')g(x')\geq 0$ for arbitrary $g$, follows directly from the definition \eqref{14}. 

\medskip\noindent
{\em Products of observables}

The operator representation of $X$ and $P$ allows the introduction of a product structure for observables that differs from the classical product. This ''quantum product'' is induced by the operator product. For example, the observable $X^2$ is associated to the squared operator $\hat X^2$, i.e. 
\be\label{18}
\kl X^2\kr=\int dx x^2\rho(x,x).
\ee
It can be realized as a classical observable by
\be\label{19}
(X^2)_\tau=\int dx f^*_\tau(x)x^2f_\tau(x),
\ee
such that the expectation value \eqref{18} obeys also the classical statistical rule
\be\label{20}
\kl X^2\kr=\sum_\tau p_\tau(X^2)_\tau.
\ee
More formally, the quantum product $X\circ X\to X^2$ can be defined as a map relating the classical  observables $X$ and $X^2$. The particular choice \eqref{19} represents an equivalence class of similar products that all result in the same quantum operator product \cite{CWAA}.

In contrast, the classical product $A\cdot B$ of two observables $A$ and $B$ is given by $(A\cdot B)_\tau=A_\tau B_\tau$, such that $(X\cdot X)_\tau=X^2_\tau$ differs from $(X^2)_\tau$, with 
\be\label{21}
\kl X\cdot X\kr=\sum_\tau p_\tau X^2_\tau=\sum_\tau p_\tau\big(\int dx f^*_\tau(x)xf_\tau (x)\big)^2.
\ee
We observe that the expectation value of the classical product $\kl X\cdot X\kr$ cannot be computed from the information contained in $\rho(x,x')$. It is therefore not compatible with the coarse graining, in contrast to the quantum product. 

The difference between the quantum and classical products may be visualized by comparing the quantum dispersion $\Delta^2_x$ with the classical dispersion $(\Delta^{(cl)}_x)^2$, 
\be\label{22}
\Delta^2_x=\kl X^2\kr-\kl X\kr^2~,~(\Delta^{(cl)}_x)^2=\kl X\cdot X\kr-\kl X\kr^2,
\ee
for an ensemble with $\kl X\kr=0$. Consider a sequence $\tau$ for which the occupied bits are concentrated in a region of space around $x_p$,  while the empty bits are concentrated around $-x_p$, such that $|f_\tau|^2(x)$ remains an even  function of $x$ centered around $x_p$ and $-x_p$. This sequence does not contribute to the classical dispersion, since $X_\tau=0$ and therefore $(X_\tau)^2=0$, while it contributes to the quantum dispersion due to $(X^2)_\tau\approx x^2_p>0$. We notice that the quantum dispersion exceeds the classical dispersion, as can be seen from the identity
\be\label{23}
\Delta^2_x-(\Delta^{(cl)}_x)^2=\sum_\tau p_\tau\int dx f^*_\tau(x)(x-X_\tau)^2 f_\tau(x)\geq 0.
\ee
A detector which signals activity in a region around $x_p$ whenever there is a substantial imbalance between occupied and empty bits will measure the quantum dispersion. Only measurements with this type of detector can be described in terms of the reduced information of the coarse grained system.

The issue of commutativity or non-commutativity always refers to a particular choice of a product between two observables. We have seen that different types of products between observables can be defined consistently. The appropriate choice of the product depends on the particular setting how measurements are done \cite{CWPO}. If a subsystem can be described by a reduced amount of information, the results of ''good measurements'' of two properties of the subsystem can only concern the information characterizing the subsystem. In other words, the statistical outcome of such measurements must be computable in terms of the information available for the subsystem. In our case, the information characterizing the subsystem is coded in the density matrix $\rho$. Position measurements which are consistent with this coarse grained system have to be associated to the quantum product $X^2$, since the classical dispersion and $\kl X\cdot X\kr$ are not computable in terms of $\rho$. 

The use of the quantum product for repeated measurements of $X$ in a one particle system has a natural interpretation. We can associate the positive and normalized function
\ba\label{24}
w(x)&=&\rho(x,x)=\sum_\tau p_\tau|f_\tau(x)|^2,\nn\\
w(x)&\geq& 0~,~\int dxw(x)=1
\ea
with the probability to find the particle at the position $x$. Repeated measurements of the particle position have then the standard statistical interpretation in the coarse grained system, the expectation value for $n$ measurements being
\be\label{24A}
\kl X^n\kr=\int dx x^nw(x).
\ee
In other words, the conditional probability to find in a second measurement the particle at $x$, if it has been found at $x$ in the first measurement, equals one. The classical product $X\cdot X$ has no such simple interpretation. We could define a classical probability for simultaneous activity at $x$ and $y$ by
\be\label{24B}
w_{cl}(x,y)=\sum_\tau p_\tau|f_\tau|^2(x)|f_\tau|^2(y)
\ee
and express
\be\label{24C}
\kl X\cdot X\kr=\int dxdy~xy~w_{cl}(x,y).
\ee
This does not seem to be a very suitable one-particle concept. It is then not surprising that the classical product is not compatible with the coarse graining to a one-particle subsystem.

As long as we are interested only in observables of the type $X^n$ we only need the information about the diagonal elements $\rho(x,x)=w(x)$. The expectation values of arbitrary observables $G(X)$, defined by 
\be\label{29A}
\big( G(X)\big)_\tau=\int dx f^*_\tau(x)G(x)f_\tau(x)
\ee
can be computed in terms of $w(x)$ as
\be\label{25}
\kl G(X)\kr=\sum_\tau p_\tau\big(G(X)\big)_\tau=
\int dx w(x)G(x)=\text{tr}\big(G(\hat X)\rho\big).
\ee
Only the modulus of $f_\tau(x)$ is needed. This changes for derivative observables as the momentum $P$ or the squared momentum $P^2$ defined by
\be\label{27}
(P^2)_\tau=\int dx f^*_\tau(x)(-\partial^2_x)f_\tau(x)=
\int dx|\partial_x f_\tau(x)|^2.
\ee
(Note that $P^2$ generalizes the roughness \eqref{4} to the case of two species. We keep a one-dimensional notation with obvious generalization to $d$ dimensions.) We now need the off-diagonal elements of $\rho$ according to 
\ba\label{28}
\kl P^2\kr&=&\sum_\tau p_\tau(P^2)_\tau=\text{tr}(\hat P^2\rho)\nn\\
&=&\int dx dx'\delta(x'-x)(-\partial^2_x)\rho(x,x').
\ea
Again, only the quantum product $P^2$ is compatible with the coarse graining, while the classical product
\be\label{29}
\kl P\cdot P\kr=\sum_\tau p_\tau P^2_\tau=-\sum_\tau p_\tau\big (\int dx f^*_\tau(x)\partial_x f_\tau(x)\big)^2
\ee
is not computable in terms of $\rho$. We conclude that the coarse grained system is described by ``incomplete statistics'' \cite{3}: The classical product of two observables is not available for the subsystem.

These observations extend to products of position and momentum observables. While the classical product $X\cdot P$ is not available for the coarse grained system, the quantum products defined by
\ba\label{30}
(XP)_\tau&=&\int dx f^*_\tau(x) x(-i\partial_x)f_\tau(x),\nn\\
(PX)_\tau&=&\int dx f^*_\tau(x)(-i\partial_x)x f_\tau(x),
\ea
obey
\be\label{31}
\kl XP\kr=\text{tr}(\hat X\hat P\rho)~,~\kl PX\kr=\text{tr}(\hat P\hat X\rho).
\ee
The quantum product is non-commutative
\be\label{32}
XP-PX=i.
\ee
Thus our classical statistical ensemble implements Heisenberg's uncertainty relation for measurements of position and momentum which are compatible with the coarse graining. For such measurements the quantum product appears directly in the measurement correlation for a sequence of measurements of position and momentum. This correlation has to be based on appropriate conditional probabilities and differs from the classical correlation which would be based on the classical product $X\cdot P$. The measurement correlation for two measurements, one of position and the other of momentum, is given \cite{CWAA, CWPO} by 
\be\label{33}
\kl XP\kr_m=\frac12(\kl XP\kr+\kl PX\kr).
\ee
Finally, in the three dimensional case a definition of an angular momentum observable which is compatible with the coarse graining must be based on the quantum product $L_k=\epsilon_{klj}X_lP_j$ rather than involving the classical product $X_l\cdot P_k$. We conclude that all features of quantum mechanical observables are recovered for our classical statistical ensemble. Planck's quantum $\hbar$ can be induced by changing the units of $P$. 

For an appropriate time evolution of the classical probability distribution $\{p_\tau\}$ the induced time evolution of the density matrix obeys the von-Neumann equation \cite{CWPO}. This can be achieved by a unitary evolution of $\{p_\tau\}$ (rotations of the unit vector $q_\tau$ with $p_\tau=q^2_\tau$) \cite{CWT} which is compatible with the coarse graining \cite{CWQP}. The detailed dynamics can be formulated as a fundamental equation for the time evolution of $\{p_\tau\}$. In turn, for a fixed time evolution of $\{p_\tau\}$ the particular Hamiltonian governing the time evolution of the quantum subsystem can be computed. We will not discuss this issue in the present note and refer to refs. \cite{CWPO,CWT}. 

\medskip\noindent
{\em Pure quantum states}

At this point a few comments concerning the role of pure quantum states are in order. (i) The coarse graining is based on the density matrix \eqref{14} while the expectation value of the function observable $\kl f(x)\kr$ given by eq. \eqref{3} plays no role. In general, the expectation values of $X$ or $P$ cannot be computed from $\kl f(x)\kr$. A pure quantum state obeys
\be\label{34}
\int dx' \rho(x,x')\rho(x',x'')=\rho(x,x'')
\ee
and allows the introduction of a quantum wave function $\psi(x)$ obeying
\be\label{35}
\rho(x,x')=\psi(x)\psi^*(x')
\ee
in the usual way. In general, $\psi(x)$ differs from $\kl f(x)\kr$. 

(ii) An exception is given for a pure classical state for which $p_\tau=0$ except for one particular $\bar\tau$ with $p_{\bar\tau}=1$. In this case on has
\be\label{36}
\psi(x)=\kl f(x)\kr=f_{\bar\tau}(x).
\ee
One may imagine a deterministic time evolution which remains within the space of pure classical states - either in a discrete way by ``jumping'' between different sequences $\tau$, or by extending the space of states by replacing the discrete labels $2n_\alpha-1=\pm 1$ by $\cos\varphi_\alpha$ and performing continuous changes of the angles $\varphi_\alpha$. Such a time evolution realizes a unitary evolution of $\psi(x)$ since by definition of $f_\tau(x)$ the norm is preserved. This special case can be viewed as a deterministic hidden variable theory for quantum mechanics, with discrete or continuous hidden variables given by $n_{\alpha,s}$ or $\cos\varphi_{\alpha,s}$. 

(iii) The coarse graining leading to the density matrix \eqref{14} can be performed in two steps. In a first step we consider an intermediate statistical ensemble with states characterized by functions $\bar f_\alpha(x_i)$. For this purpose we collect all bit sequences $\tau$ which lead to $f_{\alpha,\tau}(x_i)$ within a given function-interval $\bar f_\alpha(x_i)-\gamma_\alpha\leq f_{\alpha,\tau}(x_i)\leq\bar f_\alpha(x_i)+\gamma_\alpha$. The function intervals are chosen such that an arbitrary $f_{\alpha,\tau}(x_i)$ belongs precisely to one of the intervals labeled by $\bar f_\alpha(x_i)$. (This should hold for all space intervals labeled by $x_i$.) We may define a collective or coarse grained probability
\be\label{37}
p\big[\bar f_\alpha(x_i)\big]=\sum_\tau p_\tau|_{\bar f_\alpha(x_i)},
\ee
where the sum extends over all $\tau$ for which $f_{\alpha,\tau}(x_i)$ belongs to the function-interval labeled by $\bar f_\alpha(x_i)$. In the continuum limit $p\big[\bar f_\alpha(x)\big]$ becomes a functional of $f_\alpha(x)$. Many different sequences $\tau$ can ''belong'' to the same function $\bar f_\alpha(x_i)$ such that the mapping $\{p_\tau\}\to p\big [\bar f _\alpha(x)\big]$ is not injective and constitutes indeed a first coarse graining step. The second step defines $\rho(x,x')$ (for $F=2$ and complex $f$) as
\be\label{38}
\rho(x,x')=\sum_{\bar f(x_i)}p\big[\bar f(x_i)\big]\bar f(x_i)\bar f^*(x'_i),
\ee
where the sum extends over all function-intervals of the complex function $\bar f(x_i)$. This equals the definition \eqref{14}. In the continuum limit $p\big[\bar f(x)\big]$ is a functional of the complex function $\bar f(x)$. Again, the map $p\big[\bar f(x)\big]\to \rho(x,x')$ is not injective. In the language of ref. \cite{CWPO} we can identify the collective states labeled by $\bar f(x)$ with the ''microstates'' $\sigma$, while $\tau$ labels the ''substates''.

(iv) We can define a pure state on the level of the first coarse graining step by $p\big[\hat f(x_i)\big]=1$ for some selected function $\hat f$, while $p\big[\bar f(x_i)\big]=0$ for all other $\bar f$. In this case one finds a pure state quantum density matrix with associated wave function
\ba\label{39}
\psi(x)=\hat f(x).
\ea
We may view $p[\bar f]$ as the probability to find a given possible wave function $\bar f$. In turn, for a pure quantum state one particular $\hat f=\psi$ is selected with unit probability. Mixed quantum states arise if $p[\bar f]$ differs from zero for two or more functions $\bar f$. Inversely, it is a necessary condition for a pure quantum state that only one $\hat f$ contributes in eq. \eqref{38}. If one is interested only in position and momentum observables and their quantum products one could forget about the basic states $\tau$ and start directly with a classical statistical ensemble characterized by $p[\bar f]\geq 0,\int{\cal D}\bar fp[\bar f]=1$. A pure state could be interpreted as a deterministic hidden variable theory with hidden variables now given by the real functions $f_1(x),f_2(x)$. It is sufficient that $\hat f=\psi$ evolves in time according to a Schr\"odinger equation. This aspect shares common features with Bohm's interpretation of quantum mechanics \cite{BO} or with the setting of ref. \cite{Man}. Mixed quantum states find an interpretation in terms of ``classical probabilities'' $p[\bar f]$ according to eq. \eqref{38}.

(v) One could be tempted to use use at the intermediate level a classical statistical  interpretation for which both the position and the momentum observables $X$ and $P$ are realized as classical observables. In this case they would have a ''sharp''value $\bar X[\bar f],\bar P[\bar f]$ for every state $\bar f(x)$, given by eqs. \eqref{11}, \eqref{12} with $f_\tau$ replaced by $\bar f$. However, for a deterministic hidden variable theory of this type the explanation of a non-vanishing dispersion $\Delta^2_x$ is difficult to understand. We will also see that simple local projection observables for a quantum  particle cannot be implemented in such a setting. 

If we start from the microphysical ensemble with states $\tau$ the status of the observables $X$ and $P$ for a given microstate $\bar f(x)$ is different. They are now ''probabilistic observables'' \cite{CWPO,PO} with a probability distribution of possible measurement values for every given $\bar f(x)$. Then $\bar X[\bar f]$ and $\bar P[\bar f]$ denote the means obtained from this distribution. The ''quantum product'' $X^2,P^2,XP$ etc. differs again from the ''classical product'' at the intermediate level, which reads $(\overline{A\times B})[\bar f]=\bar A[\bar f]\bar B[\bar f]$. Also the classical product at the intermediate level is not compatible with the coarse graining \eqref{38} (except for pure states). For example, $X\times X$ involves four powers of $\bar f$ and $\kl X\times X\kr$ cannot be computed from the information contained in $\rho(x,x')$. This contrasts with the quantum product $X^2$. The above arguments concerning the relevance of the quantum product for position measurements remain valid. Again, the central ingredient for the appearance of non-commutativity in a classical statistical ensemble is the presence of a product structure for observables that differs from the classical product and is appropriate for the measurements of position and momentum.

\medskip\noindent
{\em Local projectors}

In the second part of this note we address another crucial feature of quantum  physics, namely the postulate that individual measurements of an observable always result in eigenvalues of the associated operator. This is particularly striking for operators with a discrete spectrum as angular momentum. Every individual measurement can only yield one of its discrete eigenvalues. We want to show here how this postulate can follow from the standard classical statistical interpretation of measurements within our setting of an ensemble based on states specified by bit sequences $\tau$. 

Finding classical probabilities which reproduce for some given observables the same expectation values as in quantum mechanics is per se not difficult. For example, one may associate the states of the ensemble with $\bar f$, where the variables $\bar f$ or ``function variables'' $\bar f(x)$ correspond to the real and imaginary parts of a quantum wave function. Classical observables take for each $\bar f$ the same expression as quantum observables in terms of the wave function. For any probability distribution $p[\bar f]$ the quantum expectation values will then be found according to the density matrix \eqref{38}. (Classical ensembles with less degrees of freedom have also been found - see, for example, ref. \cite{Coh} for integer spins.) However, this is not enough for an explanation of measurements and correlations in quantum mechanics. For example, for three-state or ``spin-one'' quantum mechanics a spin observable would take in every state $\bar f$ the value corresponding to the expectation value in the associated pure quantum state. According to classical statistics, measurements should then find a continuity of values between $-1$ and $1$, rather then the discrete values $-1,0,1$ predicted by quantum mechanics.

In contrast, the discreteness of quantum physics can be realized in a straightforward way if a classical observable associated to a quantum operator with a discrete spectrum has for every state $\tau$ one of the discrete values belonging to its spectrum. According to the basic setting of classical statistics each measurement finds then a value within the spectrum, while the probability distribution $\{p_\tau\}$ specifies with which probability a given eigenvalue of the associated operator will be found. We demonstrate this issue by discussing the classical statistical implementation of two familiar quantum operators with a discrete spectrum, local projectors and the spin.

Within quantum mechanics, we may define ''local projectors'' or ''interval observables'' which multiply the wave function by a factor
\be\label{40}
J(\bar x,a)=\theta\left (\bar x+\frac a2-x\right )\theta\left 
(x-\bar x+\frac a2\right ).
\ee
They have the property that $J(\bar x,a)\psi(x)=\psi(x)$ for $x\in  I(\bar x,a)$, and $J(\bar x,a)\psi(x)=0$ otherwise, with $I(\bar x,a)$ the interval $\bar x-\frac a2<x\leq \bar x+\frac a2$. The corresponding operators are projectors, $J^2(\bar x,a)=J(\bar x,a)$. Therefore the eigenvalues of the operator $J(\bar x,a)$ are $1$ or $0$, and the interpretation is simple: either a particle detector covering the interval $I(\bar x,a)$ finds a particle $(J=1)$ or not $(J=0)$. The eigenstates with eigenvalue $J=1$ are all $\psi(x)$ which vanish identically outside the interval $I(\bar x,a)$, while eigenstates with $J=0$ vanish inside the interval. We will use the discrete formulation of functions as used for $f_\tau(x_i)$ in eq. \eqref{6}. We concentrate on intervals $I(x_i,\epsilon)=I(x_i)$ of size $\epsilon$ and on interval observables $J(x_i)=J(x_i,\epsilon)$. The eigenfunctions $\psi_{(x_i)}(x)$ of the interval observables $J(x_i)$ form a basis, such that arbitrary $\psi(x)$ can be written as $\psi(x)=\sum_{x_i}a_{(x_i)}\psi_{(x_i)}(x)$. They equal one inside the interval $I(x_i)$ and vanish outside, with an additional factor $\epsilon^{-1/2}$ in the continuum normalization. The functions $\psi_{(x_i)}(x)$ are also eigenfunctions of the position operator $\hat X$, with eigenvalue $x_i$. We can write
\be\label{40A}
\hat X=\sum_ix_iJ(x_i).
\ee

So far standard quantum mechanics. In order to realize the interval observables $J(x_i)$ as classical observables we need to associate to every sequence $\tau$ a number $\big (J(x_i)\big)_\tau=1,0$. With respect to each $J(x_i)$ the states $\tau$ decay into two classes, depending if $\big(J(x_i)\big)_\tau$ equals one or zero. On the coarse grained level the expectation value of $J(x_i)$ reads
\be\label{41}
\kl J(x_i)\kr=\rho(x_i,x_i)=w(x_i).
\ee
Consistency therefore requires that the assignments of $\big(J(x_i)\big)_\tau$ and probability distributions $\{p_\tau\}$ obey
\be\label{42}
\kl J(x_i)\kr=\sum_\tau p_\tau\big(J(x_i)\big)_\tau=\sum_\tau 
p_\tau f^*_\tau (x_i)f_\tau(x_i).
\ee
We define here $f_\tau(x_i)$ by eq. \eqref{6}. The normalization \eqref{7} imples $|f_\tau(x_i)|^2\leq 1$ and therefore $0\leq J(x_i)\leq 1$, as it should be. However, if we admit arbitrary sequences $\tau$ it seems difficult to find an assignment $\big(J(x_i)\big)_\tau$ such that eq. \eqref{42} holds for arbitrary probability distributions $\{p_\tau\}$. 

At this point we recall that the quantum wave function $\psi(x)$ is supposed to describe a one-particle-state. In contrast, arbitrary probability distributions $\{p_\tau\}$ describe states with an arbitrary particle number. We therefore have to select the one-particle states. They correspond to a particular class of probability distributions $\{p_\tau\}$. As a first example we consider ''locally concentrated sequences'' for which $f_\tau(x_i)$ vanishes except for one particular $\hat x_i(\tau)$, reflecting a ``particle number'' one at $\hat x_i$, and zero elsewhere. For locally concentrated sequences the normalization of $f_\tau$ implies $|f_\tau(\hat x_i(\tau)|^2=1$. It is then straightforward to define the classical interval observables as 
\be\label{43}
\big(J(x_i)\big)_\tau=|f_\tau(x_i)|^2=0,1.
\ee
In this case the possible values of measurements of $J(x_i)$ are indeed given by the values $(0,1)$ of the classical observables. They equal, in turn, the eigenvalues of the associated quantum operators. Probability distributions $\{p_\tau\}$ for which $p_\tau$ differs from zero only for the locally concentrated sequences describe one particle states. 

The notion of one particle states can be extended beyond the locally concentrated sequences. It is sufficient to associate to each sequence $\tau$ a particular space interval located at $\hat x_i(\tau)$, for example the one where $|f_\tau(x_i)|^2$ reaches its maximum. We then define the interval observables by
\be\label{43A}
\big(J(\hat x_i(\tau)\big)_\tau=1~,~\big(J(x_i\neq \hat x_i(\tau)\big)_\tau=0.
\ee
The ``allowed probability distributions'' $\{p_\tau\}$ for a one particle state are the ones which obey eq. \eqref{42}. The other ``forbidden probability distributions'' describe states with an admixture of contributions with total particle number different from one. For one particle states the total particle number observable is trivial and may be associated with $\sum_{x_i}\big(J(x_i)\big)_\tau=1$. 

The reader may notice that the notion of one particle states depends on the resolution $\epsilon$. A one particle state for resolution $\epsilon$ remains a one particle state for resolution $n\epsilon,n>1$, as follows from combining $n$ intervals of size $\epsilon$. However, a one particle state obeying eq. \eqref{42} for a resolution $\epsilon $ needs not to obey similar conditions for subintervals of $I(x_i,\epsilon)$. A one particle state with resolution $\epsilon$ may therefore contain multiparticle states for a resolution $\epsilon'<\epsilon$. This closely reflects known physical properties. A one atom state for a resolution distance exceeding sufficiently the atom size appears as a multiparticle state with nucleons and electrons on a smaller resolution scale. 

It is at the level of the projection observables that a deterministic setting based on ``hidden variables'' $\bar f(x)$ fails to account for the discreteness of quantum mechanics. While the expectation values of $X$ and $P$ can be reproduced correctly in such a setting, there seems to be no ``classical'' projection observable of the type $\big(J(x_i)\big)(\bar f)$ which associates to each state or function $\bar f(x)$ one of the two allowed values $1$ or $0$. In contrast, the microphysical ensemble with states $\tau$ allows $\big(J(x_i)\big)(\bar f)$ to be a probabilistic observable which has a probability distribution of values $1$ or $0$ for each ``microstate'' $\bar f(x)$. 

\medskip\noindent
{\em Spin}

Another characteristic discrete quantum degree of freedom is the spin of a particle. For particles with internal degrees of freedom we can consider a higher species number $F$. For a spin one half particle we take $F=4$, leading to two-component complex functions $f_\tau=(f_{1,\tau}+if_{2,\tau},f_{3,\tau}+if_{4,\tau})$. Replacing $f^*_\tau$ by $f^\dagger_\tau,f^\dagger f= f^2_1+f^2_2+f^2_3+f^2_4$, all previous formulae for position and momentum observables are easily generalized. We next define three two-level observables by
\ba\label{44}
(S_1)_\tau&=&2\theta\Big(\int dx\big(f_{1,\tau}(x)f_{3,\tau}(x)-f_{2,\tau}(x)f_{4,\tau}(x)\big)\Big)-1,\nn\\
(S_2)_\tau&=&2\theta \Big(\int dx\big(f_{1,\tau}(x)f_{4,\tau}(x)-f_{2,\tau}(x)
f_{3,\tau}(x)\big)\Big)-1,\nn\\
(S_3)_\tau&=&2\theta\Big(\int dx\big(f^2_{1,\tau}(x)+f^2_{2,\tau}(x)\nn\\
&&-f^2_{3,\tau}(x)-f^2_{4,\tau}(x)\big)\Big)-1.
\ea
In every state $\tau$ they take one of the discrete values $\pm 1$, $(S^2_k)_\tau=(S_k)^2_\tau =1$, such that measurements of this observable should only find the values $+1$ or $-1$, according to the rules of classical statistics. We will identify the observables $S_k$ with the spin of a particle in the direction $k$, up to a normalization factor $\hbar/2$. We further define local spin observables by 
\be\label{45}
\big (S_k(x_i)\big)_\tau=(S_k)_\tau\big(J(x_i)\big)_\tau.
\ee
Their measurement can yield the values $+1,-1$, or $0$.

The allowed probability distributions $\{p_\tau\}$ for one particle states with spin are restricted by the conditions
\ba\label{46}
\kl J(x_i)\kr&=&\sum_\tau p_\tau\big(J(x_i)\big)_\tau=\sum_\tau p_\tau f^\dagger_\tau(x_i)f_\tau(x_i),\nn\\
\kl S_1(x_1)\kr&=&\sum_\tau p_\tau\big (S_1(x_1)\big)_\tau=2\sum_\tau p_\tau
\big(f_{1,\tau}(x_i)f_{3,\tau}(x_i)\nn\\
&&+f_{2,\tau}(x_i)f_{4,\tau}(x_i)\big)\nn\\
\kl S_2(x_i)\kr&=&\sum_\tau p_\tau\big (S_2 (x_i)\big)_\tau=2\sum_\tau p_\tau
\big(f_{1,\tau}(x_i)f_{4,\tau}(x_i)\nn\\
&&-f_{2,\tau}(x_i)f_{3,\tau}(x_i)\big),\nn\\
\kl S_3(x_i)\kr&=&\sum_\tau p_\tau\big (S_3(x_i)\big)_\tau=\sum_\tau p_\tau
\big(f^2_{1,\tau}(x_i)+f^2_{2,\tau}(x_i)\nn\\
&&-f^2_{3,\tau}(x_i)-f^2_{4,\tau}(x_i)\big).
\ea
In the complex basis the generalization of the density matrix \eqref{14} is a complex hermitean $2\times 2$ matrix. Eq. \eqref{46} implies the standard relation
\be\label{49}
\kl S_k(x)\kr=\text{tr}\big (\hat S_k(x)\rho\big)=\text{tr}\big(\tau_k\rho(x,x)\big),
\ee
with operators for the local spin
\be\label{50}
\hat S_k(x)(y,y')=\tau_k\delta(x-y)\delta(y-y').
\ee
Similarly, the spin observables obey
\be\label{51}
\kl S_k\kr=\text{tr}(\hat S_k\rho)~,~\hat S_k(y,y')=\tau_k\delta(y-y')=\int dx\hat S_k(x).
\ee
As it should be, the product of two different spin operators is not commutative, and we can define the associated quantum product for the classical spin observables $(S_k)_\tau$. Again, the classical product of spin observables, $(S_k\cdot S_l)_\tau=(S_k)_\tau(S_l)_\tau$, cannot be computed from the information available at the coarse grained level and we encounter incomplete statistics. 

We can express the expectation values \eqref{46} in terms of the functions $\bar f_\alpha(x_i)$ defined on an intermediate level of coarse graining using eq. \eqref{37},
\ba\label{47}
&&\kl J(x_i)\kr=\int {\cal D}\bar f p[\bar f]\bar J(x_i)[\bar f],\nn\\
&&\kl S_k(x_i)\kr\int{\cal D}\bar f p[\bar f]\bar S_k(x_i)[\bar f],
\ea
with
\ba\label{48}
\bar J(x_i)[\bar f]&=&\bar f^\dagger(x_i)\bar f(x_i),\nn\\
\bar S_1(x_i)[\bar f]&=&2\big(\bar f_1(x_1)\bar f_3(x_1)+\bar f_2(x_i)\bar f_4(x_i)\big)\nn\\
&=&\bar f^\dagger(x_i)\tau_2\bar f(x_i),\nn\\
\bar S_2(x_i)[\bar f]&=&2\big(\bar f_1(x_i)\bar f_4(x_i)-\bar f_2(x_i)\bar f_3(x_i)\big)\nn\\
&=&\bar f^\dagger(x_i)\tau_2\bar f(x_i),\nn\\
\bar S_3(x_i)[\bar f]&=&\bar f^\dagger(x_i)\tau_3\bar f(x_i).
\ea
The complex two-component function $\bar f(x)$ has all the properties of the quantum wave function for a spin one half particle. In particular, we notice the identity $\bar S^2_1(x_i)+\bar S^2_2(x_i)+\bar S^2_3(x_i)=\bar J^2(x_i)$ for arbitrary $\bar f$, and the normalization of $\bar f$ related to the identity $\sum_{x_i}\bar J(x_i)=1$. 

For a microphysical classical statistical ensemble which describes an isolated particle the ``one particle condition'' \eqref{48} must be preserved by the time evolution of the probability distribution $\{p_\tau\}$. The normalization of $\bar f$ implies that such a time evolution describes a generalized rotation in the real Hilbert space spanned by the functions $\bar f_\alpha(x_i)$. If the time evolution remains compatible with the complex structure this transfers to a unitary transformation in a complex Hilbert space. The infinitesimal unitary transformations define the hermitean Hamiltonian for the quantum time evolution. (If for suitable subensembles the average values of $J(x_i),S_k(x_i)$ and a local momentum observable $P(x_i)$ obey an appropriate ``purity constraint'' \cite{CWAA,CWPO} we can actually use their mean values in the subensemble in order to {\em define} the functions $\bar f_\alpha(x_i)$ by eq. \eqref{48}. This issue will be discussed in a separate paper. An independent definition of the function observables $f_\tau(x_i)$ is no longer necessary in this setting. For the description of an isolated particle it is then sufficient that the purity constraint is preserved by the time evolution.)

At this stage we have implemented the quantum spin observable $S_k$ as classical observables that take values $+1$ or $-1$ for every classical state $\tau$. This has to be generalized for arbitrary directions of the spin observables. This generalization requires that the action of rotations of the spin can be implemented on the level of the probability distributions $\{p_\tau\}$ such that eqs. \eqref{47}, \eqref{48} hold for spins $S_\omega$ in arbitrary directions, with a suitable assignment $(S_\omega)_\tau=\pm 1$. An explicit construction how the spin rotations are implemented on the level of classical probability distributions can be found in ref. \cite{CWPO}. For continuous rotations an infinite number of states $\tau$ is required, $N_s\to\infty$, while for finite $N_s$ only a discrete subgroup of the rotations can be realized. This does not matter in practice, since even for finite $\delta$ equal to the Planck length and $V$ some atomic volume the number of states $N_s=2^{4V/\delta^3}$ is extremely high. Similar constructions can realize the angular momentum observables $L_k$ as discrete classical observables with fixed values $(L_k)_\tau$ which belong to the spectrum of the quantum operators $\hat L_k$. 

In conclusion, we have presented an explicit classical statistical ensemble for which a one-particle state can be defined for a subclass of probability distributions $\{p^{(1)}_\tau\}$. All properties of an isolated particle can be expressed in terms of the coarse grained information contained in a density matrix $\rho(x,x')$ which is computable form the probability distribution $\{p^{(1)}_\tau\}$. One-particle observables as position, momentum or spin are realized as standard classical observables in the microphysical classical ensemble. However, their classical products cannot be defined in terms of $\rho(x,x')$ - the coarse grained system is described by incomplete statistics. On the other hand, our system allows the definition of a non-commutative quantum product for these observables which remains compatible with the coarse graining. This quantum product provides for the correlations of measurements of properties of an isolated particle. We demonstrate explicitly how the quantum formalism with non-commuting operators emerges from our classical statistical description. While on the microphysical level both the classical and the quantum product of the one-particle observables can be defined, only the quantum product ``survives'' the coarse graining. In this sense the coarse graining of the information for subsystems is the origin of the non-commutative structure of quantum physics.

\end{document}